\documentclass[aps,pra,twocolumn,amsmath,amssymb,floatfix,superscriptaddress]{revtex4-2}

\usepackage{graphicx}
\usepackage[normalem]{ulem}
\usepackage{xcolor}
\usepackage[title]{appendix}
\usepackage{hyperref}
\usepackage{booktabs}
\usepackage{longtable}
\usepackage{multirow}
\usepackage{siunitx}

\makeatletter
\def\tableautorefname{Tab.\hskip\z@skip}
\def\subtableautorefname{Tab.\hskip\z@skip}
\def\figureautorefname{Fig.\hskip\z@skip}
\def\subfigureautorefname{Fig.\hskip\z@skip}
\def\sectionautorefname{Sec.\hskip\z@skip}
\def\subsectionautorefname{Sec.\hskip\z@skip}
\def\subsubsectionautorefname{Sec.\hskip\z@skip}
\def\appendixautorefname{App.\hskip\z@skip}
\def\chapterautorefname{Chapter\hskip\z@skip}

\def\equationautorefname~#1\null{%
  Eq.\hskip\z@skip~(#1)\null
}
\makeatother



\definecolor{darkblue}{HTML}{00317d}
\definecolor{lightblue}{HTML}{008dff}
\definecolor{pink}{HTML}{ff73b6}
\definecolor{purple}{HTML}{c701ff}
\definecolor{teal}{HTML}{4ecb8d}
\definecolor{orange}{HTML}{ff9d3a}
\definecolor{yellow}{HTML}{f9e858}
\definecolor{red}{HTML}{d83034}

\newcommand{\affA}{Van der Waals-Zeeman Institute, Institute of Physics,
University of Amsterdam, 1098 XH Amsterdam, Netherlands}
\newcommand{\affB}{QuSoft, Science Park 123, 1098 XG Amsterdam, the Netherlands}
\newcommand{\affC}{School of Physics, University of New South Wales, Sydney, New South Wales 2052, Australia}

\usepackage{pifont}

\newcommand{\colorsquare}[1]{\raisebox{0pt}{{\color{#1}\scalebox{0.75}{\ding{110}}}}}
\newcommand{\colorcircle}[1]{\raisebox{0pt}{{\color{#1}\scalebox{0.75}{\ding{108}}}}}

\begin{document}

\title{Long-lived metastable states in the 4f$^{13}$5d6s configuration of Yb$^+$}

\author{Z.~E.~D.~Ackerman}\affiliation{\affA}
\author{A.~Cadarso Quevedo}\affiliation{\affA}
\author{Ilango~Maran}\affiliation{\affA}
\author{L.~P.~H.~Gallagher}\affiliation{\affA}
\author{R.~J.~C.~Spreeuw}\affiliation{\affA}
\author{J.~C.~Berengut}\affiliation{\affC}
\author{R.~Gerritsma}\affiliation{\affA}\affiliation{\affB}

\newcommand{\MM}[1]{{\color{Orange}{#1}}}
\newcommand{\RG}[1]{{\color{Blue}{#1}}}
\newcommand{\RS}[1]{{\color{Red}{#1}}}
\newcommand{\CP}[1]{{\color{Violet}{#1}}}
\newcommand{\ND}[1]{{\color{Green}{#1}}}

\date{\today}

\begin{abstract}

We study the occurrence of long-lived metastable states in the 4f$^{13}$5d6s electron configuration of Yb$^+$. By optical pumping of a single trapped ion on the $^2F^\text{o}_{7/2}\rightarrow (7/2,0)_{7/2}$ transition at 377.5~nm, we prepare a wide range of metastable  electronic states. We use a co-trapped control ion to sympathetically cool the  spectroscopy ion, allowing us to accurately time its subsequent decay. We record a strong decay signal corresponding to a lifetime of 0.92(8)~s, a weaker decay signal with lifetime 9.8(+2.9, -2.0)~s, and find evidence for a much longer lifetime, $>$~30~s. We identify the metastable states with these lifetimes qualitatively, and corroborate our results with atomic structure calculations that support the observed lifetimes and decay paths. These long-lived states provide new opportunities in qubit and qudit state detection and optical clocks.

\end{abstract}

\maketitle

\emph{Introduction.}---
Metastable states in trapped ions play a crucial role in quantum information \cite{Schindler:2013,Allcock:2021,Ransford:2021,Yang:2022}, optical atomic clocks~\cite{Ludlow:2015,Huntemann:2016,Hausser:2024} and tests of fundamental physics \cite{Peik:2004,dreissen_improved_2022,Hur:2022,door2024searchnewbosonsytterbium,Wilzewski:2025}. The ion $^{171}$Yb$^+$ stands out as the species of choice to combine experimentally accessible laser cooling transitions with the smallest nonzero nuclear  spin~\cite{Olmschenk:2007}. Its level structure thus offers first-order magnetic field insensitivity on transitions resulting in unmatched qubit coherence times~\cite{Wang:2021} while minimizing complications due to hyperfine structure. 
In Yb$^+$, three metastable states are widely used for clocks and quantum technology: the doublet $5^2D_{3/2}$ and $5^2D_{5/2}$ with lifetimes of $\sim$~50~ and $\sim$~7~ms, respectively~\cite{Yu:2000,Taylor:1997}, and the extremely long-lived $4^2F^\text{o}_{7/2}$ state with a lifetime of $\sim$~3.2~years~\cite{Lange:2021,Lange:2021v2}. The comparatively short lifetime of the $^2D$ states makes them less attractive for qubit manipulation and atomic clocks~\cite{Ludlow:2015}. At the same time, the tiny coupling strength on the $6 ^2S_{1/2}\rightarrow4^2F^\text{o}_{7/2}$ octupole transition~\cite{Lange:2021} complicates its use in quantum computing. A metastable state with an intermediate lifetime and coupling strength would be an attractive alternative for quantum computing and clocks.

Atomic structure calculations reveal metastable states besides the ones mentioned above in the 4f$^{13}$5d6s electron configuration~\cite{Fawcett:1991,Biemont:1998,Safronova:2009,Porsev:2012}. Most prominently, Fawcett and Wilson calculated a lifetime of 5.2~s for the state ${}^3[3/2]^\text{o}_{5/2}$~\cite{Fawcett:1991}. The 4f$^{13}$5d6s configuration also produces electronic states with angular momentum $j>7/2$, for which electric dipole decay is strictly forbidden. The longevity of these states has not been tested in experiment.

In this paper, we prepare a wide range of metastable states in the 4f$^{13}$5d6s electron configuration, and measure their lifetime on the second scale. In order to accurately time the decay, we use a co-trapped control ion to cool and reveal the presence of the spectroscopy ion while it remains in a metastable state. We find a strong decay signal corresponding to a lifetime of 0.92(8)~s and a weaker decay signal with lifetime 9.8(+2.9, -2.0)~s. Furthermore, we find evidence of a much longer decay time $>$~30~s. We analyze the data and the decay paths of the various product states qualitatively and perform atomic structure calculations with the AMBiT package~\cite{Kahl:2019}, corroborating our results. We conclude that the 0.92(8)~s lifetime belongs to the state ${}^3[3/2]^\text{o}_{5/2}$. The longer lifetimes may be attributed to terms with angular momentum $j>7/2$, namely ${}^3[11/2]^\text{o}_{9/2}$ and ${}^3[7/2]^\text{o}_{9/2}$. Finally, we investigate the possible application of the studied states and transitions in quantum technology.

\begin{figure*}
    \centering
   \includegraphics[width=1\linewidth]{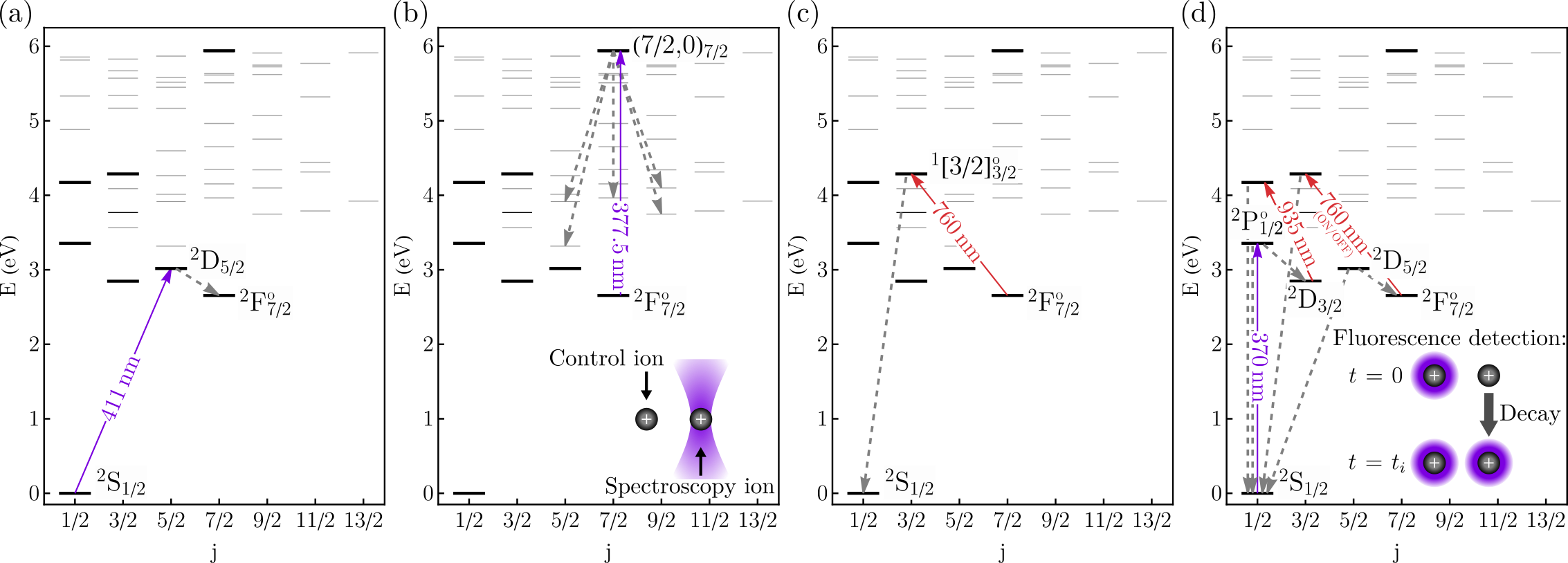}
    \caption{Experimental sequence and level scheme. (a) Both ions are prepared in the $^2F^\text{o}_{7/2}$ state using a 411~nm laser pulse driving the $^2S_{1/2}\rightarrow\,^2D_{5/2}$ transition followed by decay to $^2F^\text{o}_{7/2}$. (b) The spectroscopy ion is driven on the 377.5~nm $^2F^\text{o}_{7/2}\rightarrow (7/2,0)_{7/2}$ transition, followed by possible decay to metastable states. (c) The control ion is returned to the ground state via the 760~nm $^2F^\text{o}_{7/2}\rightarrow {}^1[3/2]^\text{o}_{3/2}$ transition. (d) Fluorescence detection with lasers at 370~nm and 935~nm to observe at what time the spectroscopy ion decays back to a state belonging to the fluorescence cycle ($^2S_{1/2}$ and $^2D_{3/2}$). We can perform this last step with or without the 760~nm laser, allowing us to distinguish decay paths including or excluding decay to the $^2F^\text{o}_{7/2}$ state. The control ion keeps the spectroscopy ion cold while it is in a metastable state, such that we can accurately determine its return time.}
    \label{fig_sequence}
\end{figure*}

\emph{Setup.}---We trap two $^{174}$Yb$^+$ ions in a microfabricated 3-dimensional linear Paul trap. The trap is driven by an rf field at 7.985~MHz, resulting in secular trap frequencies of $\omega_{x,y,z}\simeq2\pi~\times~$(500, 500, 280)~kHz. Excess micromotion was compensated to amplitudes $\leq$~100~nm. The ion trap is installed in an ultra-high vacuum chamber with a background pressure of 3.0(4)$\times10^{-8}$~Pa. The ions are Doppler cooled on the $^2S_{1/2}\rightarrow \,^2P^\text{o}_{1/2}$ transition using a laser beam at 370~nm, which is aligned 45\textdegree~to each trap axis. From the $^2P^\text{o}_{1/2}$ state, $\sim$~0.5~\% decays to the metastable $^2D_{3/2}$ state. A 935~nm laser beam is used to repump ions from the $^2D_{3/2}$ state via the ${}^3[3/2]^\text{o}_{1/2}$ state back into the cooling cycle. The 370~nm and the 935~nm laser beam are on continuously during the experiment. We image the ions by collecting their fluorescence light at 370~nm onto an EMCCD camera. Permanent magnets are used to generate a magnetic field of $\sim$~0.5~mT in the $z$-direction.

The experimental sequence is shown in Fig.~\ref{fig_sequence}. We prepare the ions in the long-lived $^2F^\text{o}_{7/2}$ state by driving the $^2S_{1/2}\rightarrow\,^2D_{5/2}$ quadrupole transition at 411~nm (Fig~\ref{fig_sequence}a). The 411~nm beam is counter-propagating to the direction of the 370 nm beam and its polarization and laser propagation direction are at 45\textdegree~to the magnetic field. Continued driving for 100~ms accumulates all population in the $^2F^\text{o}_{7/2}$ state~\cite{Feldker:2020} via the decay $^2D_{5/2}\rightarrow{}^2F^\text{o}_{7/2}$. Next, we use a focused beam to drive the spectroscopy ion on the electric dipole (E1) transition $^2F^\text{o}_{7/2}\rightarrow (7/2,0)_{7/2}$ at 377.5~nm for 50~ms (Fig~\ref{fig_sequence}b). From there, the ion can follow a number of decay paths, passing a number of potential metastable states that we discuss below.

After the 377.5~nm pulse, we empty the $^2F^\text{o}_{7/2}$ state with a 100~ms pulse of 760~nm light (Fig~\ref{fig_sequence}c). This returns the control ion to the laser cooling cycle with near unit probability. The 760~nm polarization and laser propagation direction are at 45\textdegree~and 90\textdegree~to the magnetic field, respectively, to address all magnetic sub-states of $^2F^\text{o}_{7/2}$~\cite{Roos:2000}.

Observing the fluorescence signal on a CCD camera, we can detect the presence of the dark spectroscopy ion by the position of the control ion. As soon as the spectroscopy ion decays to a state that is part of the fluorescence cycle, it will start to emit light again and we record the return time with $\sim$~15~ms accuracy (Fig~\ref{fig_sequence}d). 

We implement two variations of this scheme, in which we either keep on or switch off the 760~nm $^2F^\text{o}_{7/2}$ clear-out laser during waiting. Decay paths that end in the state $^2D_{3/2}$ lead to the decayed ion starting to fluoresce again. Decays directly to $^2F^\text{o}_{7/2}$ can only be observed in the experiment where we switch on the 760~nm laser during waiting. Decay to the state $^2D_{5/2}$ is followed by  further decay to $^2S_{1/2}$ and $^2F^\text{o}_{7/2}$ with a decay time of $\sim 7$~ms and the 83~\% branching fraction in favor of the $^2F^\text{o}_{7/2}$ state. We also perform control measurements for each variation in which we omit the 377.5~nm pulse to check that all other laser operations work according to expectation. We avoid systematic drifts by continuous cycling between the four measurement settings.

Occasionally, the focused laser makes the control ion dark instead of the spectroscopy ion. This may be due to a background gas collision immediately after the pulse. Furthermore, it can be the case that both ions become dark. Both types of event are removed from the data as we cannot be certain what happened. These occurrences can be caused by crosstalk in the addressed beam and amount to 4.8(5)~\% and 7.9(8)~\% of the total data, respectively. We limit the wait time to 30~s to assure a reasonable data collection rate while avoiding limitations due to background gas collisions, and load new ions in case the spectroscopy ion remained dark.

We perform additional measurements to rule out effects of background gas collisions that may quench metastable states within the observed time frame. Furthermore, we repeated the experimental sequences with varied laser power of the 370~nm and 935~nm beams used for Doppler cooling and fluorescence detection to assure that the lifetimes are not limited by off-resonant scattering. More details on these additional measurements can be found in the appendix.

\begin{figure}[h!]
    \centering
   \includegraphics[width=0.95\linewidth]{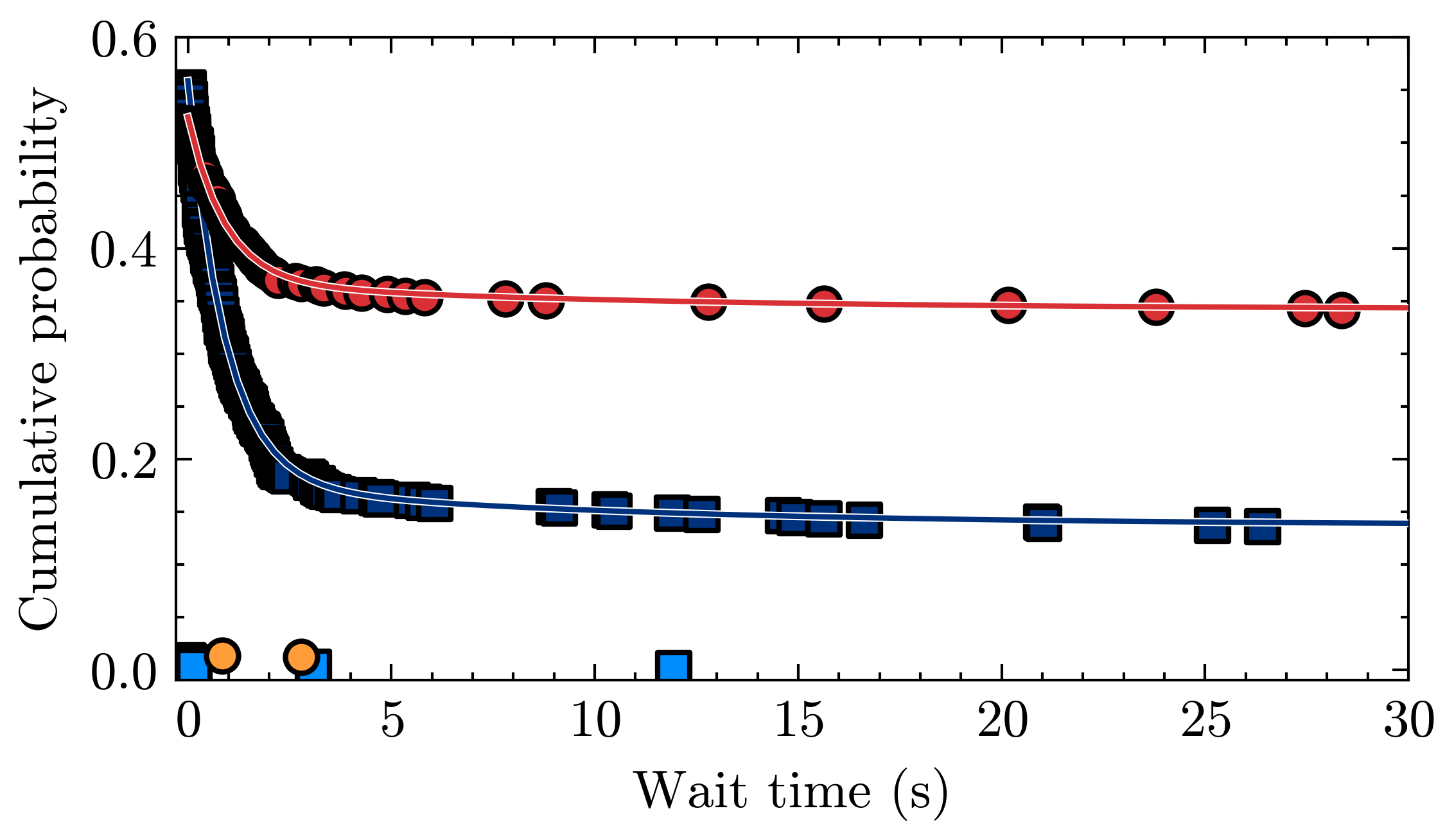}
    \caption{Measured cumulative probability that the spectroscopy ion is dark as a function of the wait time, for different experimental sequence configurations. (\colorsquare{darkblue}) The 760~nm laser on during the wait time. (\colorcircle{red}) The 760~nm laser off during the wait time. (\colorsquare{lightblue}) Control measurement in which the 377.5~nm pulse is omitted with the 760~nm laser on. (\colorcircle{orange}) Control measurement with the 760~nm laser off. The blue (red) curve represents the function $C(t)=-\alpha P(t)+\beta$, as explained in the text, fitted to the data with the 760~nm laser on(off) during the wait time.}
    \label{probabilies}
\end{figure}

\emph{Results.}\label{results}---We perform 1000 repetitions of the experimental sequences, and record return times as well as cases in which the ion returned before $t=0$~s and cases in which the ion did not return after $t=30$~s. The results are shown in Fig.~\ref{probabilies}, where we plot the probability that the spectroscopy ion remains dark as a function of waiting time. The  probabilities with the 760~nm laser on and off are not independent because the former contains all possible decays to the states $^2D_{3/2}$, $^2D_{5/2}$ and $^2F^\text{o}_{7/2}$ and the latter decay to $^2D_{3/2}$ and 17\% of the decay to $^2D_{5/2}$.

We first fit the observed return times in the interval $t\in(0,30)$~s by performing a maximum likelihood estimation. We find that the model must allow for, at a minimum, a biexponential decay, which is described by the function $p(t)$: the probability density of ion decay at time $t$. Motivated by the overlap of the cumulative probabilities and the similar shape of the two curves in Fig.~\ref{probabilies}, we assume the same lifetimes $\tau_1$ and $\tau_2$ for the two datasets, but we allow their relative amplitude $r$ to differ between them. At the maximum likelihood, the dominant decay has a lifetime of $\tau_1=$~0.92(8)~s and the other lifetime is $\tau_2=$~9.8(+2.9, -2.0)~s, with relative amplitudes of $r_\text{ON}=$~0.91(3) and $r_\text{OFF}=$~0.86(6), with the subscript indicating the fit to the return times with the 760~nm laser ON/OFF. The statistical errors indicate where the likelihood dropped by a factor $e^{-1}$.

Next, we fit the biexponential decay to the complete dataset, including the cases where the spectroscopy ion became bright immediately and the cases where it did not return. We use the function $C(t)=-\alpha P(t)+\beta$, with $P(t)=\int_0^t p(t')dt'$, $\beta$ the proportion of cases where the spectroscopy ion did not return, and $\alpha + \beta$ the proportion of cases where the spectroscopy ion did not become bright immediately. We fit $\alpha_\text{ON}=0.4218(19)$, $\beta_\text{ON}=0.1371(9)$, $\alpha_\text{OFF}=0.1823(14)$ and $\beta_\text{OFF}=0.3423(7)$. We see in the data that even when the 760~nm laser is on, 13.58(13)~\% of the ions, consistent with $\beta_\text{ON}$, do not return to the fluorescence cycle within 30~s, which indicates the presence of another state with a very long lifetime. As this lifetime outlives the average time between background gas collisions as described in the appendix, we cannot reliably record it in our system.

\begin{figure}
    \centering
   \includegraphics[width=1\linewidth]{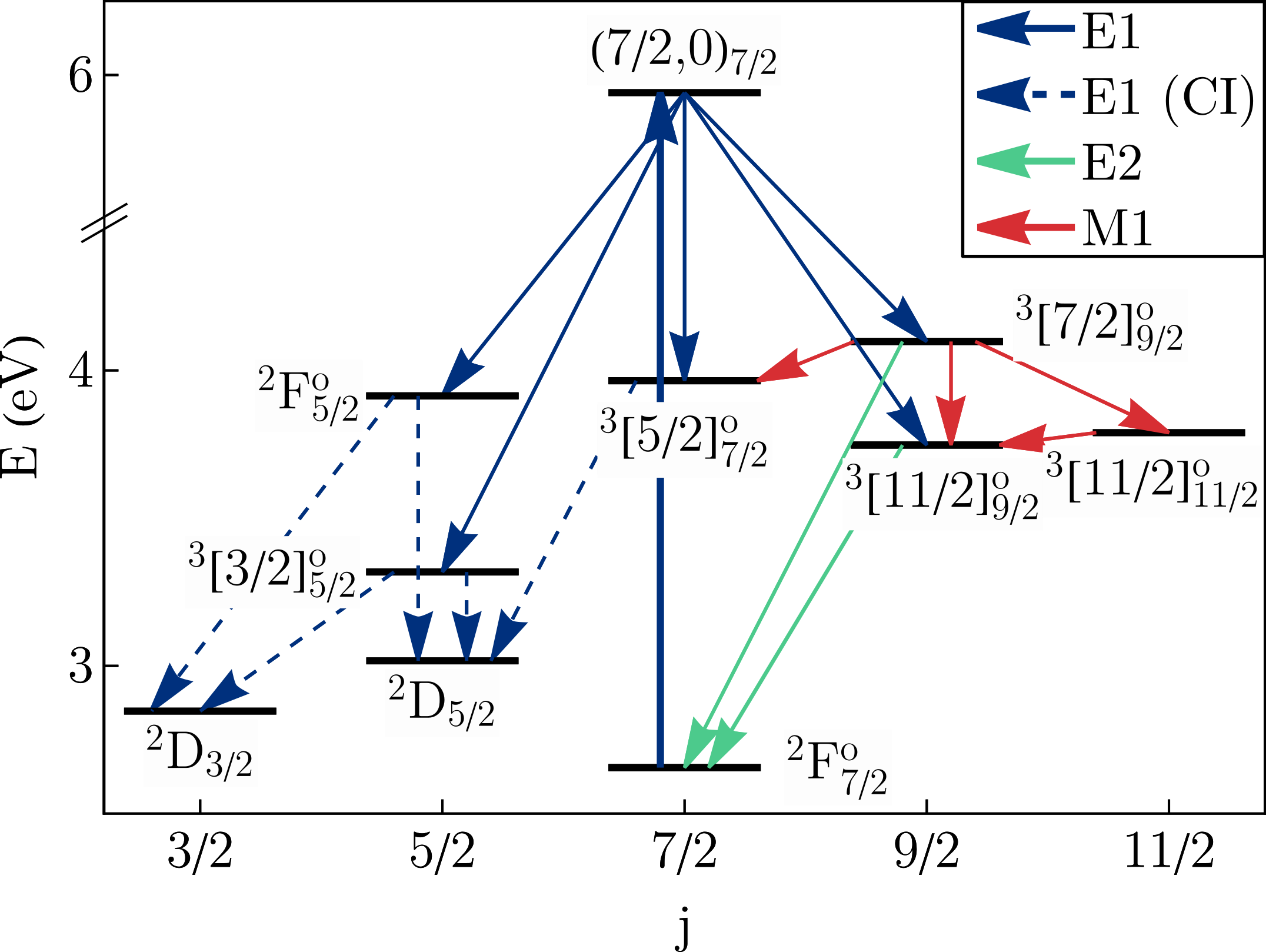}
    \caption{ Partial level scheme of Yb$^+$ showing the laser driving the $^2F^\text{o}_{7/2}\rightarrow (7/2,0)_{7/2}$ transition (thick blue arrow). We include the dominant E1 decay channels observed in discharge experiments~\cite{Meggers:1967} as well as the weaker configuration interaction-induced (CI) E1 decay channels of the metatsable states, M1 decay and E2 decay. To improve clarity we only indicate expected dominant decay channels for the metastable states of interest. Data taken from the NIST database~\cite{NIST_DATABASE}. }
    \label{levelscheme}
\end{figure}

\emph{Identification of metastable states.}---While our scheme allows to probe a large number of possible metastable states, assigning the measured decay rates to particular states is a challenging task. In the following we identify some possible candidates.
 
Experimental data obtained in discharge experiments~\cite{Meggers:1967}, as well as theoretical predictions based on pseudo-relativistic Hartree-Fock methods~\cite{Fawcett:1991} indicate that the decay of $(7/2,0)_{7/2}$ is dominated by transitions to the lowest lying terms in the 4f$^{13}$5d6s configuration, which are shown in Fig.~\ref{levelscheme}. Besides the driven 377.5~nm line, Meggers~\cite{Meggers:1967} reported four strong decay lines to the terms ${}^3[3/2]^\text{o}_{5/2}$, ${}^3[5/2]^\text{o}_{7/2}$, ${}^3[11/2]^\text{o}_{9/2}$ and ${}^3[7/2]^\text{o}_{9/2}$ at 3.32, 3.96, 3.75~eV and 4.10~eV, respectively; with a weaker decay line to the term $^2F^\text{o}_{5/2}$ of 4f$^{13}$6s$^2$ at 3.91~eV. The calculations of Fawcett and Wilson~\cite{Fawcett:1991} produced the same lines and added smaller decay channels to ${}^3[5/2]^\text{o}_{5/2}$ and ${}^3[9/2]^\text{o}_{7/2}$ at 4.01 and 4.35~eV, respectively (not shown in Fig.~\ref{levelscheme}). 

We now analyze the possible decay paths of these states qualitatively. The longevity of ${}^3[3/2]^\text{o}_{5/2}$(3.32~eV) was noted by Fawcett and Wilson, who calculated a lifetime of 5.2~s~\cite{Fawcett:1991}. The long lifetime can be explained by a number of factors, namely the small energy gap to the $^2D$ states it decays to and the reliance on configuration interaction to open the otherwise dipole-forbidden 4f$^{13}$5d6s~$\rightarrow$~4f$^{14}$5d decay channel. Moreover, the term has a strong spin quartet $^4P^\text{o}_{5/2}$ character (in $LS$-coupling), suppressing decay to the doublet states by spin selection rules. 

Fawcett and Wilson also calculated lifetimes of 36.8 and 19.7~ms for ${}^3[5/2]^\text{o}_{5/2}$(4.01~eV) and ${}^3[5/2]^\text{o}_{7/2}$(3.96~eV), respectively. These lifetimes would be too short to detect in our experiment, as they are shorter than the time required to faithfully repump the control ion, (Fig~\ref{fig_sequence}c). Nevertheless, the decay timescale would correspond to a failure to create a dark spectroscopy ion, evidence of which can be seen in our data at $t=0$.. 

The ${}^3[11/2]^\text{o}_{9/2}$(3.75~eV) and ${}^3[7/2]^\text{o}_{9/2}$(4.10~eV) terms cannot dipole decay to the $^2D$ states due to rigorous selection rules. Their decay may populate the $^2F^\text{o}_{7/2}$ state via an electric quadrupole (E2) transition. The ${}^3[7/2]^\text{o}_{9/2}$(4.10~eV) state may in addition  magnetic dipole (M1) decay to  ${}^3[5/2]^\text{o}_{7/2}$(3.96~eV), ${}^3[11/2]^\text{o}_{9/2}$(3.75~eV) and ${}^3[11/2]^\text{o}_{11/2}$(3.79~eV). The very recent work of Ref.~\cite{McMillin:2026} estimates a M1 limited lifetime of $\sim$~9.9~s for ${}^3[7/2]^\text{o}_{9/2}$(4.10~eV), when neglecting configuration interactions. A recent measurement found a lifetime of 0.589(53) ms for the $^2F^\text{o}_{5/2}$ state~\cite{sidorov2026}, which is similarly too short to observe in our experiment.

\emph{Atomic structure calculations}\label{atomic_structure_calculations}---We now present atomic structure calculation performed with AMBiT~\cite{Kahl:2019} in support of our experimental results. AMBiT treats the valence electrons and holes using configuration interaction, and incorporates the effects of core-valence correlations using many-body perturbation theory (CI+MBPT). 

\begin{table}[h]
\centering
\caption{Lifetimes (in seconds unless specified) of a few low energy and metastable states of Yb$^+$. Column 2 presents values calculated using AMBiT. Columns 3 and 4 lists values theoretically predicted by Fawcett~\cite{Fawcett:1991} and experimentally measured (if available), respectively. Details on the choice of parameters and settings for the calculation are given in the appendix.}
\renewcommand{\arraystretch}{1.3} 
\begin{tabular}{lccc}
\hline\hline
State & $\tau_{\rm calc}$  & $\tau_{\rm Faw}$~\cite{Fawcett:1991} & $\tau_{\rm Exp}$ \\
\hline
$^2{P}_{1/2}^\text{o}$ & 6.2 ns & 8.8 ns & 8.12(2) ns~\cite{Olmschenk:2009}\\

$^2{D}_{3/2}$ & 44.3 ms &40 ms &  52.7(2.4) ms~\cite{Yu:2000}\\

$^2{D}_{5/2}$ & 5.8 ms & 7.1 ms & 7.2(3) ms~\cite{Taylor:1997} \\

$^2{F}_{7/2}^\text{o}$ & 4.7$\times10^7$ &1.3$\times10^8$ & 9.96(50)$\times10^7$\cite{Lange:2021,Lange:2021v2}\\

\hline
$^3{[3/2]}_{5/2}^\text{o}$ & 3.1 & 5.2 & -\\
$^2{F}_{5/2}^\text{o}$ & 0.52 ms & -&  0.589(53) ms~\cite{sidorov2026}\\
$^3{[5/2]}_{7/2}^\text{o}$  &13.1 ms &19.7 ms& -\\
$^3{[11/2]}_{9/2}^\text{o}$ &2456 &-& -\\
$^3{[7/2]}_{9/2}^\text{o}$ & 19.8& - & -\\
\hline\hline
\end{tabular}
\label{Tab:Lifetimes}
\end{table}

The results can be seen in Table.~\ref{Tab:Lifetimes}, where we report lifetimes of a few states calculated with AMBiT. We compare the calculations with other experimental results and the calculations of Fawcett and Wilson. We see that for the known lifetimes, our results fall within the same order of magnitude. For the metastable states studied here, we find three lifetimes in the second range or above, belonging to the states ${}^3[3/2]^\text{o}_{5/2}$(3.32~eV), ${}^3[11/2]^\text{o}_{9/2}$(3.75~eV) and ${}^3[7/2]^\text{o}_{9/2}$(4.10~eV).

 Based on the above analysis we connect $\tau_1$ to ${}^3[3/2]^\text{o}_{5/2}$(3.32~eV). Assuming CI-induced E1 decay to the $^2D$ states dominates, we find a branching fraction of $\sim$~71(4)~\% in favor of the $^2D_{5/2}$ state. The lifetime $\tau_2$ fits a small amount of data ($\sim$~3~\% of the total data). A possible explanation for its appearance in both datasets with 760~nm laser on and off could be the term ${}^3[7/2]^\text{o}_{9/2}$(4.10~eV), that can M1 decay to ${}^3[5/2]^\text{o}_{7/2}$(3.96~eV) followed by E1 decay to $^2D_{5/2}$~\cite{McMillin:2026}. An identification of the long decay time $>30$~s with the state ${}^3[11/2]^\text{o}_{9/2}$(3.75~eV) is plausible~\cite{McMillin:2026}. 

\emph{Applications.}---Metastable states are used for qubit state detection in trapped ion quantum computing~\cite{Sauter:1986} and offer further benefits in a so-called {\it omg} scheme (short for optical-frequency, metastable-state, and ground-state qubits)~\cite{Allcock:2021,Ransford:2021,Yang:2022,McMillin:2026}. The metastable states in the 4f$^{13}$5d6s electron configuration discussed in this paper offer promising candidates for both qubit and qudit state detection in Yb$^+$.

Within this electron configuration, the state ${}^3[3/2]^\text{o}_{5/2}$(3.32~eV) provides an attractive alternative for electron shelving. Due to its intermediate lifetime and coupling strength, it would be a more reliable and faster option compared to the $^2D$ and $^2F^\text{o}_{7/2}$ states, respectively. Qudit-based quantum computing using the isotope $^{173}$Yb$^+$ may especially benefit, as a large number of shelving operations are needed for qudit readout~\cite{Lanyon:2008,Campbell:2014,Senko:2015,Ringbauer:2022}.

The state may be reached from the ground state via the magnetic quadrupole (M2) transition $^2S_{1/2}\rightarrow {}^3[3/2]^\text{o}_{5/2}$ at 373.7~nm. Our AMBiT calculations give an M2 linestrength of $S\sim 27\times (\alpha_\text{FS}/2)^2$ in atomic units, with $\alpha_\text{FS}$ the fine-structure constant. Hyperfine enhancements of the $^2S_{1/2}\rightarrow {}^3[3/2]^\text{o}_{5/2}$ coupling may be expected for both $^{173}$Yb$^+$ and $^{171}$Yb$^+$~\cite{Dzuba:2016}. To ensure rapid return to the ground state $^2S_{1/2}$ after shelving, repumping of the ${}^3[3/2]^\text{o}_{5/2}$ state is best done on an M1 or E2 transition. Several of such transitions are available in the visible spectrum, coupling to terms in the 4f$^{13}$5d6s and 4f$^{13}$5d$^2$ configurations.  

Finally, we note that in neutral Yb a similar inner-shell orbital clock transition was observed~\cite{Ishiyama:2023,ishiyama:2025} recently with applications in clocks and precision spectroscopy~\cite{Dzuba:2018}.

\vspace{2mm}
We thank the group of Florian Schreck for making available the Sr 689~nm laser for wavelength calibration. We thank Nella Diepeveen and Jook Walraven for fruitful discussions and Martijn Reitsma for helping set up the AMBiT calculations. This work was supported by the Dutch Research Council (Grant Nos. 680.91.120, VI.C.202.051, 680.92.18.05 and OCENW.M.22.403).

\emph{Data availability.}---Data for this article including raw experimental data and data files to reproduce the figures are available at \href{https://doi.org/10.21942/uva.31647100.v1}{https://doi.org/10.21942/uva.31647100.v1}.

%

\onecolumngrid
\vspace{1cm}

\twocolumngrid
\section*{End matter}

\begin{figure}[h!]
    \centering
\includegraphics[width=1\linewidth]{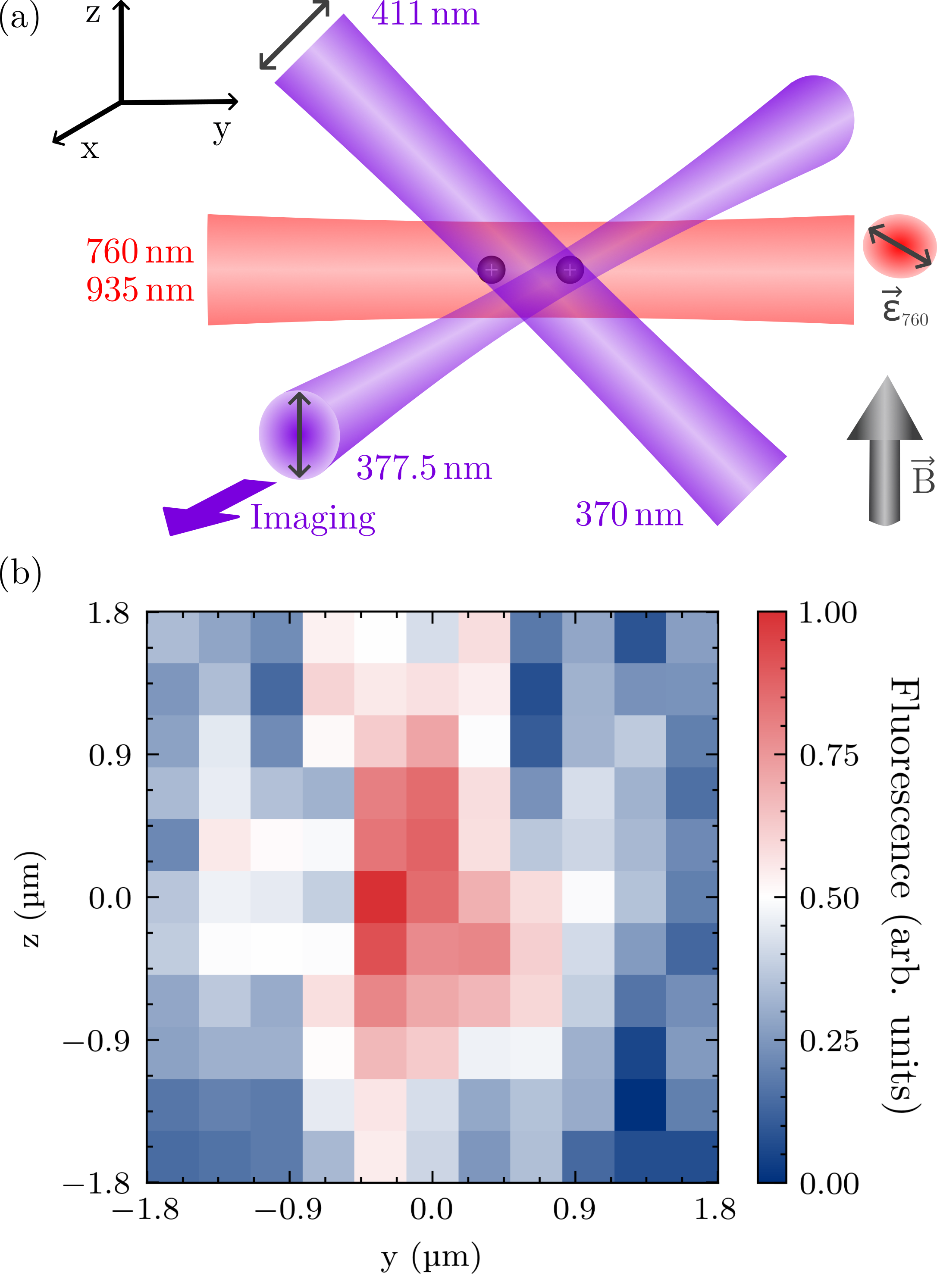}
    \caption{(a) Laser configuration for spectroscopy. (b) Ion fluorescence as a function of 377.4~nm beam position.}
    \label{geometries_and_beammap}
\end{figure}

\emph{Appendix A: Focused laser alignment.}---The laser geometries used for spectroscopy are shown in Fig.~\ref{geometries_and_beammap}a. The 377.5~nm laser beam used to drive the $^2F^\text{o}_{7/2}\rightarrow (7/2,0)_{7/2}$ transition is tightly focused by an aspheric lens with a high effective numerical aperture ($\sim$~0.45), which is also used for imaging. The 377.5~nm laser beam is aligned on a single ion by tuning its frequency to the $^2D_{5/2}\rightarrow {}^3[5/2]^\text{o}_{3/2}$ transition at 377.4~nm. Via the ${}^3[5/2]^\text{o}_{3/2}$ state, the ion returns to the fluorescence cycle, making the 377.4~nm a repump transition. The transition frequency we measured in $^{174}$Yb$^+$ is 794.449501(30) THz, with uncertainty that is limited to the uncertainty quoted from our wavemeter (High Finesse WS8-10). The laser beam is steered using a spatial light modulator while the 411~nm laser continuously couples the ion to the $^2D_{5/2}$ state. The recorded ion fluorescence as a function of the beam position allows mapping of the light field in the imaging plane of the ion. The resulting 2D map is shown in Fig.~\ref{geometries_and_beammap}b. We find beam waists $w_z=0.94(8)$~{\textmu}m and $w_y=1.31(12)$~{\textmu}m, sufficiently small compared to the inter-ion distance of $\sim$~8~{\textmu}m. The estimated beam center is overlapped with the spectroscopy ion before starting the experimental sequence. The $^2F^\text{o}_{7/2}\rightarrow (7/2,0)_{7/2}$ transition is found by observing ions going dark over $\sim$~3~GHz around $\sim$~ 794.257~THz.


\emph{Appendix B: Background gas collisions.}---Background gas collisions can quench metastable states, or vice versa, populate them. Moreover, chemical reactions may occur leading to the formation of molecular ions that we cannot distinguish from long-lived electronic states without time-consuming mass spectrometry.  The background gas in an ultrahigh vacuum environment is comprised mostly of H$_2$ gas. Formation of YbH$^+$ after collisions has been studied in Ref.~\cite{Hoang:2020}. 

Both chemical reactions and state quenching require short-range (Langevin) collisions between the ion and the background gas molecule. We measure the Langevin collision rate using the techniques developed in Refs.~\cite{ZaehringerPhD:2012,FuerstPhD:2019,Aikyo:2020,Hahn:2025}. We co-trap one $^{176}$Yb$^+$ ion with one $^{174}$Yb$^+$. The latter is laser cooled and fluoresces in the process while the former remains dark. Following a background gas collision with sufficient energy, there is a 50\% chance of the positions of the bright and dark ion interchanging. By recording these position changes over a long period of time, we obtain a lower limit of the background gas Langevin collision rate of 1/(43(3)~s). Our lifetime measurements are limited by this rate, i.e. we cannot reliably determine decay rates that are slower than this collision rate. 
  
To check our assumption that all Langevin collisions result in ion crystal melting, we repeat the measurement with $\omega_r=2\pi~\times~$800~kHz, corresponding to a $\sim$ three times higher energy barrier for ion position exchange~\cite{Aikyo:2020,Hahn:2025}. We find a consistent ion reordering rate of 1/(46(4)~s). We rule out radiation pressure effects by confirming that the two-ion crystal does not have a preferred orientation. 


\begin{figure}[h!]
    \centering
   \includegraphics[width=0.95\linewidth]{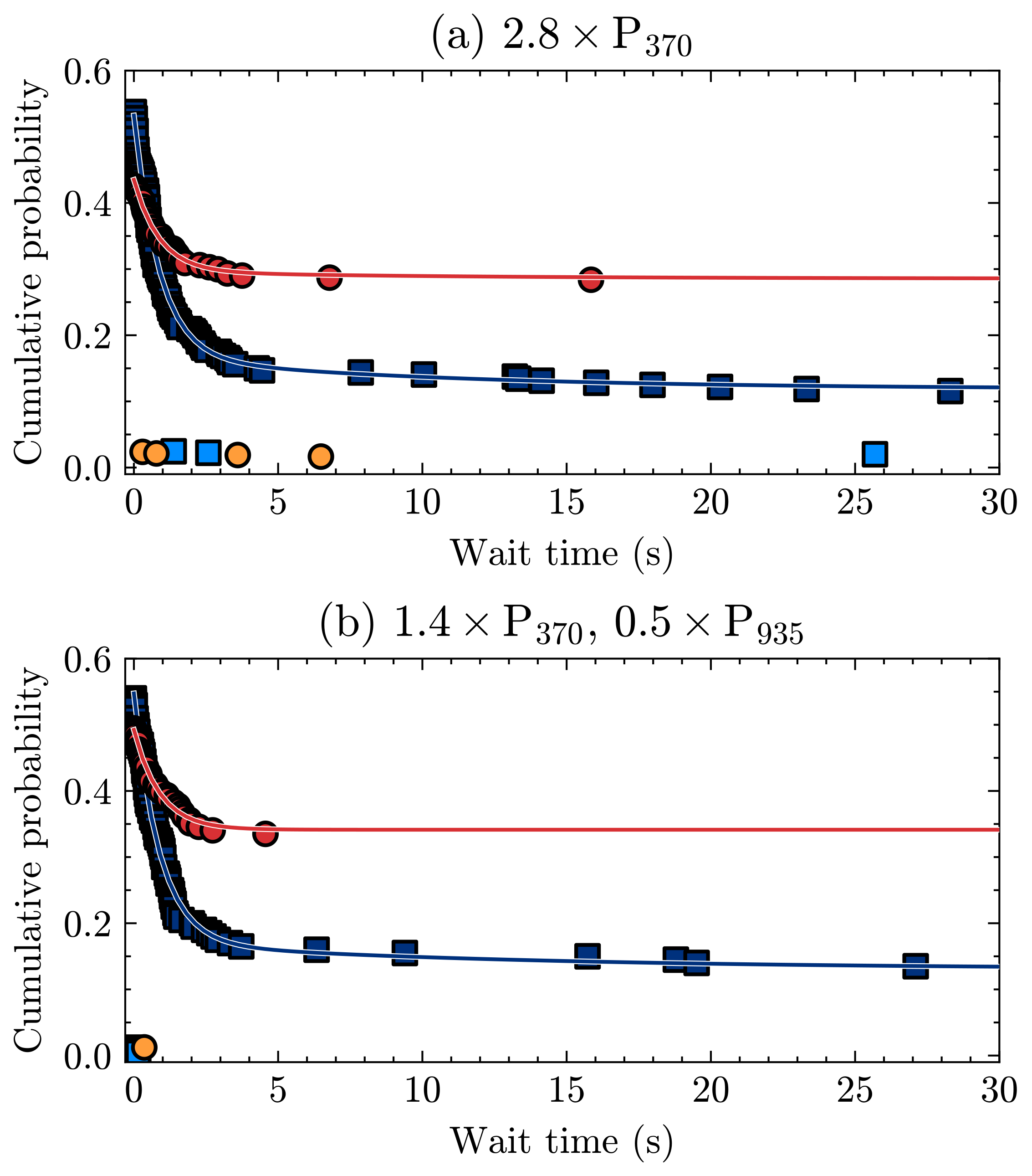}
    \caption{Measured cumulative probability that the spectroscopy ion is dark as a function of the wait time, for different experimental sequence configurations. (\colorsquare{darkblue}) The 760~nm laser on during the wait time. (\colorcircle{red}) The 760~nm laser off during the wait time. (\colorsquare{lightblue}) Control measurement in which the 377.5~nm pulse is omitted with the 760~nm laser on. (\colorcircle{orange}) Control measurement with the 760~nm laser off. The blue(red) curve represents the function $C(t)=-\alpha P(t)+\beta$, as explained in the text, fitted to the data with the 760~nm laser on (off) during the wait time. Compared to the laser powers used in the original measurement, (a) the 370~nm power is increased by a factor $\sim$~2.8 and (b) the 370~nm power is increased by a factor $\sim$~1.4 and the 935~nm power is halved.}
    \label{probabilies_alternative_powers}
\end{figure}

\emph{Appendix C: Effects of off-resonant scattering.}---To accurately determine the return times, the 370 nm and the 935 nm laser were kept on continuously during the experimental sequence. To investigate the influence of these lasers on the measured lifetimes, we performed two alternative sets of measurements for which we varied these laser powers, to compare these to the original measurement of which the results are described in the main text. In the first set, the 370~nm power is increased by a factor $\sim$~2.8 compared to the power $\mathrm{P}_{370}$ used for the original set. In the second set, the 370~nm power is increased by a factor $\sim$~1.4 compared to the original set, and the original 935~nm power $\mathrm{P}_{935}$ is halved. The experimental sequences are repeated 500 times and 270 times, respectively. The results are shown in figure \autoref{probabilies_alternative_powers}.

We fit the data following the same procedure as in the main text. For the first measurement set, shown in figure \autoref{probabilies_alternative_powers}a, we find at the maximum likelihood, that the dominant decay has a lifetime of $\tau_1=$~0.89(11)~s and the other lifetime is $\tau_2=$~10.2(+4.8, -2.9)~s, with relative amplitudes of $r=$~0.93(+5, -9) with the 760 nm laser off and $r=$~0.88(+4, -5) with the 760 nm laser on. We fit $\alpha_\text{ON}=0.414(4)$, $\beta_\text{ON}=0.1186(19)$, $\alpha_\text{OFF}=0.1496(21)$ and $\beta_\text{OFF}=0.2854(11)$. For the second measurement set, shown in figure \autoref{probabilies_alternative_powers}b, we find the dominant decay has a lifetime of $\tau_1=$~0.89(13)~s and the other lifetime is $\tau_2=$~12(+11, -5)~s, with relative amplitudes of $r=$~1.00(+0, -6) with the 760 nm laser off and $r=$~0.90(+5, -6) with the 760 nm laser on. We fit $\alpha_\text{ON}=0.414(5)$, $\beta_\text{ON}=0.1313(26)$, $\alpha_\text{OFF}=0.170(3)$ and $\beta_\text{OFF}=0.3214(21)$. We find both lifetimes $\tau_1$ and $\tau_2$ are, within statistical errors, the same for all three laser power configurations. The contribution of $\tau_2$ with the 760 nm laser on also remained the same, although we did see a reduction of the contribution of $\tau_2$ with the 760~nm laser off and the 935~nm power halved. We conclude that off-resonant laser scattering of the 370~nm and the 935~nm lasers does not significantly influence the reported lifetimes. Further analysis is limited by the number of measurement repetitions.


\emph{Appendix D: Atomic structure calculations.}---The atomic structure calculations in support of our experimental results have been performed with AMBiT~\cite{Kahl:2019}, an implementation of the particle-hole CI+MBPT formalism~\cite{berengut16pra} which is a natural extension of the CI+MBPT method~\cite{dzuba96pra} to account for valence holes. Here we present details specific to our calculations; further methodological details are presented in the aforementioned papers and references therein.

In the case of Yb$^+$, with $N=69$ electrons, we begin our calculation with a Dirac-Hartree-Fock (DHF) calculation in the $V^{N-2}$ approximation, including all closed-shell core orbitals up to 5s, 5p, 4d, and including 13 out of 14 electrons of the 4f shell by scaling the closed-shell DHF potential. Because the 4f orbitals change so significantly between levels with closed or open 4f shell, we find that this gives a better inital approximation, albeit at the cost of including subtraction diagrams in MBPT. In the same potential we generate a valence basis for CI that includes orbitals up to 7spdf (i.e. $n\leq7$ and angular momentum $l\leq3$), and a virtual basis up to 30spdfgh that we use for MBPT.

Our CI calculation includes all configurations that can be generated with a maximum of a single additional 4f excitation and two other valence excitations from a list of leading configurations that includes 4f$^{14}$6s, 4f$^{14}$5d, 4f$^{14}$6p, 4f$^{13}$6s5d, 4f$^{13}$6s$^2$, 4f$^{13}$6p$^2$ and 4f$^{13}$5d$^2$. Core shells up to 5s, 5p, and 4d are considered frozen at the level of CI. The resulting CI matrices have dimension $\sim$~20000 -- 30000 for each parity (even or odd) and angular momentum $J$.

Core–valence correlation effects are included using second-order MBPT by modifying the one and two-electron Slater integrals in CI and adding an effective three-body interaction. The calculations reproduce the experimental fine structure within each leading configuration reasonably well, but give a large energy gap between the configurations with closed and open 4f shells. This is improved by optimizing a constant offset $\Delta$ to the MBPT energy denominators~\cite{berengut08jpb}
\[
\frac{1}{E-E_M} \rightarrow \frac{1}{E-E_M + \Delta} ,
\]
where $E$ and $E_M$ are orbital approximations to the valence and intermediate MBPT energies. As noted in~\cite{kozlov99os}, such an energy denominator can be justified on rather general grounds because it restores the correct asymptotic behaviour in the Brillouin-Wagner perturbation theory for a large number of particles. In this work we treat $\Delta$ as a free parameter, which allows us to converge the level energies towards the experimental level structure.

The energies of different states and linestrengths of the necessary transitions are calculated with $\Delta$ values of -2.4 (P1) and -2.5 (P2). We also perform calculations with more parameter sets, P3 (which expands the valence basis to $n=8$ and $l=3$ and choosing $\Delta = -2.4$) and P4 (which adds the 4f$^{14}$5f, 4f$^{13}$6s6p and 4f$^{13}$6p5d configurations to the leading configuration set used before and $\Delta = -2.4$). P5 adds Breit interactions in the DHF calculations, while P6 takes into account the random phase approximation (see e.g.~\cite{Dzuba:2018a} and references therein) for dealing with the effect of core polarizability on the calculated linestrengths. Both P5 and P6 use the rest of the parameters as in P1. We find variations in energy at the $\sim$~0.1~eV level between the different parameter sets. The calculated linestrengths vary at the $\lesssim$~10~\% level across P1-P5, and several 10~\% when comparing to P6. Nevertheless, given the complexity of the Yb$^+$ spectrum and the limitations of our attempts to model it, we expect an error of around 20--30\% in the calculated rates.

We calculate the lifetimes and branching of the decays of each state using the expressions for the spontaneous emission rates $R$ in atomic units~\cite{Dzuba:2018}:
\begin{align}
R_\text{E1,M1}= & \frac{4}{3}\left(\alpha_\text{FS}\omega\right)^3\frac{S_\text{E1,M1}}{2j+1}\\
R_\text{E2,M2}= & \frac{1}{15}\left(\alpha_\text{FS}\omega\right)^5\frac{S_\text{E2,M2}}{2j+1}\\
R_\text{E3,M3}= & \,0.001 69\left(\alpha_\text{FS}\omega\right)^7\frac{S_\text{E3,M3}}{2j+1}
\end{align}
\noindent Here, $j$ denotes the angular momentum of the upper state and $\omega$ is the frequency of the transition in atomic units. Note that here, the linestrengths include a factor $(\alpha_\text{FS}/2)^2$ for the magnetic transitions M1, M2 and M3. We use the experimental energy differences and calculated linestrengths using parameter set P1 to obtain lifetimes for the states of interest. 

\end{document}